# Ultrafast nonlinear pulse propagation dynamics in metal-dielectric periodic photonic architectures

*Jitendra Nath Acharyya[1], Akhilesh Kumar Mishra[2*], D. Narayana Rao[3], Ajit Kumar[1], and G. Vijaya Prakash[1*]*

[1]*Nanophotonics Lab, Department of Physics, Indian Institute of Technology Delhi, New Delhi-110016, India*

[2]*Department of Physics, Indian Institute of Technology Roorkee, Roorkee-247667, India*

[3]*School of Physics, University of Hyderabad, Hyderabad-500046, India*

**Abstract:** One-dimensional (1D) metal-dielectric (MD) periodic structures take advantage of large refractive index contrast between metal and dielectrics to invoke extremely high nonlinear ultrafast responses of metal. These structures are also special due to their extremely high laser damage threshold. The Bragg like 1D MD structure $(Ag/SiO_2)_4$ enables strong optical field confinement with much enhanced nonlinear features as compared to simple metal or single $(Ag/SiO_2)_1$ structure. In the present work, the ultrafast nonlinear optical responses of the above structures are investigated via femtosecond broadband optical pump-probe technique. The enhanced nonlinear optical absorption is of reverse saturation of absorption (RSA) nature, resulted due to free-carrier absorption (FCA) and excited-state absorption (ESA) processes. The spectral nonlinearities are closely related to the pump-induced modification of the metal's dielectric functions, which are qualitatively visualized by transfer matrix and two-temperature models. The ultrafast temporal evolution of nonlinear absorption clearly demonstrated enhanced optical nonlinearity, disentangled by the electron-electron and electron-phonon dynamic interactions at picosecond time scales. A phenomenological pulse propagation model is employed that incorporates the experimentally obtained nonlinear absorption coefficients and different nonlinear effects exhibited by the system. Nonlinearity plays a crucial role in controlling the ultrafast pulse propagation and could open a new window for many nonlinear device applications. The findings of these new optical materials could possibly pave the way for promising applications in ultrafast photonics.

[*] Corresponding Author : akhilesh.mishra@ph.iitr.ac.in (AKM); prakash@physics.iitd.ac.in (GVP)

## 1. Introduction

The interaction of ultrashort laser pulses with a metallic system opens up a new field of study- the non-equilibrium dynamics of laser-heated electrons in metals. [1-3] A promising approach to modulate the metal's dielectric function is achieved by exploiting the nonlinearity of the metal via intense ultrashort laser pulses interaction. Irradiating metal nanostructures with intense ultrafast laser pulse leads to a significant increase in electron temperature compared to that in bulk medium. Hot electron gas undergoes several fast (~0.5-5 ps) elementary processes such as electron-electron (*e-e*) and electron-phonon (*e-ph*) interactions before a new equilibrium with the host lattices is reached.[1,4-6] One-dimensional (1D) metal-dielectric (MD) multilayer wavelength-ordered structures have become promising candidates to observe such enhancement in nonlinearities in a fundamental way.[4,5] As a consequence of the Bragg-like resonances, MD photonic multilayer structure is advantageous for accessing the ultrafast nonlinearities at the transmissive photonic *minibands* in the resonant and non-resonant spectral regimes.[7-10] These tailor-made photonic metal structures thus lead to the transparency in the visible and infrared regions with more than 50% transparency and show many promising applications in nano sensing, waveguiding, and all-optical switching, exploiting the tunable Kerr nonlinearity of metals.[7,8,11-14]

The ultrafast dynamics and the higher-order nonlinearities of MD metamaterials are dominated by the contribution of metal's non-instantaneous Kerr nonlinearity[15-19] and several numerical studies have been carried out in 1D and cylindrical type 2D geometries for the specific spectral region of interest.[16,17] Especially, the numerical realization of the ultrafast pulse propagation in wavelength-ordered multilayer MD structure is not straight forward because of a large variety of spatial and temporal effects, ranging from self-focusing and self-phase modulation to the formation of solitons. [20,21] The dynamics of ultrashort pulses sensitively depend upon the structural details and nonlinearities of the multilayer stack.[22-24] Hence, ultrafast pulse propagation in the presence of enhanced optical nonlinearity within such complex structures demands an in-depth investigation for a better understanding of the underlying physics.

In the present work, we investigate the ultrafast absorption dynamics and the nonlinear optical responses of one-dimensional metal-dielectric photonic structure (($Ag/SiO_2$)$_4$) realized from four bilayers of silver (Ag) of 38 nm thickness and half-wave ($\lambda/2$) thickness of fused silica ($SiO_2$). The detailed carrier dynamics involving electron-electron and electron-phonon interactions are resolved by broadband transient absorption measurements using femtosecond optical pump-probe spectroscopy, while the third-order nonlinear responses were obtained by nonlinear transmission measurements, which uses single-beam femtosecond Z-scan technique. The studies are also performed on $(Ag/SiO_2)_1$ single bilayer and bare Ag thin film. The detailed investigation of the pump-induced changes is modelled using the modified transfer matrix method (TMM) for linear and nonlinear responses. Furthermore, a simplified phenomenological pulse propagation model is employed for the ultrashort pulse propagation in such photonic structures.

## 2. Results and Discussions

## 2.1. Linear Optical Responses

Metal-dielectric multilayer structures show completely different optical responses as compared to those in bulk metals. Due to alternative wavelength-ordered 1D stack arrangement, the transmission window, known as photonic bandgap, can be tailored selectively with more than 50% transmission in the visible region. Figure 1a depicts the schematic representation of a one-dimensional metal-dielectric photonic crystal $((Ag/SiO_2)_4)$ composed of four bilayers of Ag (38 nm) and $SiO_2$ (179 nm). The detailed photonic band structure information can be visualized from linear transmission and reflection spectra, which turns out to be in good agreement with the theoretical simulations performed using the transfer matrix method (TMM).[25] The simulated energy ($E$) versus in-plane momentum vector ($k = (2\pi/d) \sin\theta$, where $d$ is the thickness of ($Ag+SiO_2$)) maps are shown in Figures 1b and 1d for transmission and reflection geometries, respectively. The small white circles in the E-$k$ maps represent the experimental data at the peak/dip corresponding to angle-resolved transmission and reflection spectra. The proposed $(Ag/SiO_2)_4$ MD structure shows photonic stopband between 2.0-3.1 eV and on both sides, a series of transmission/reflection peaks appear, which are known as photonic *minibands*. The formation of *minibands* results from a series of coupled Fabry–Pérot cavities (formed between any two consecutive metal interfaces).[7,26,27] Such minibands are not being observed in the $(Ag/SiO_2)_1$ single bilayer and single Ag thin film (Figure S1, Supporting Information).[4,7] While the higher energy (3.2 to 4.0 eV) photonic minibands are dominated by the onset of strong Ag interband transitions, whereas the lower energy minibands (1.7 to 2.0 eV) are the coupled-cavity pure photonic modes.[7] In the present study, our focus is on the nonlinear optical modulations and the ultrafast dynamics of the lower energy photonic minibands.

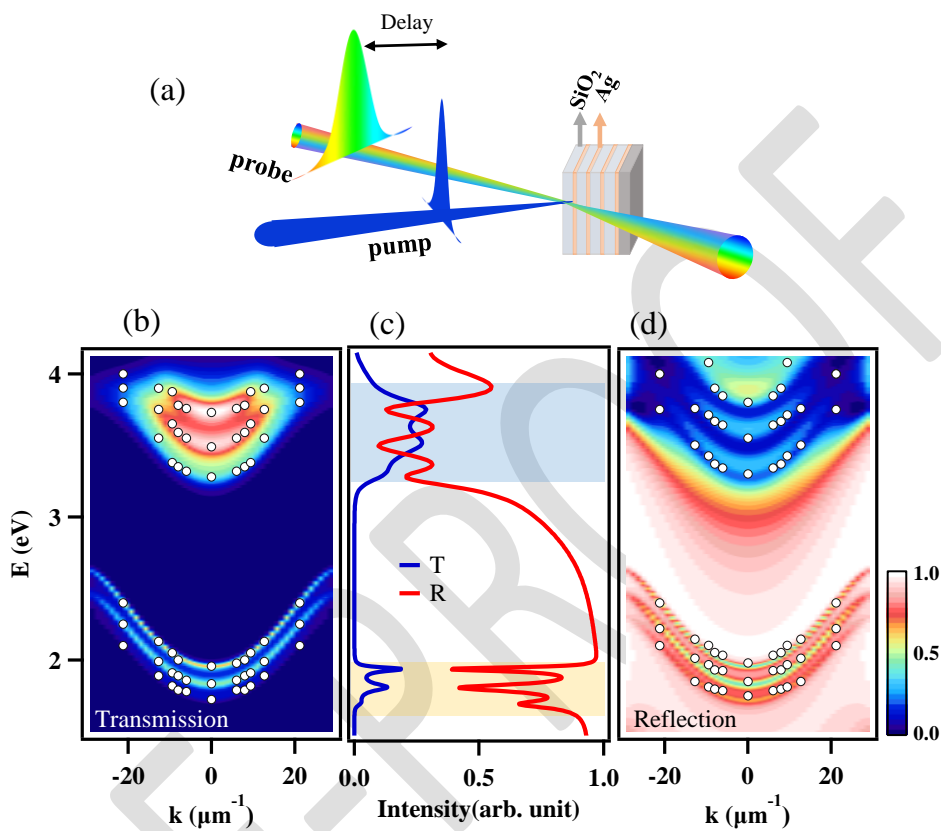

**Figure 1.** (**a**) Schematic representation of one-dimensional metal-dielectric photonic structure $(Ag/SiO_2)_4$ composed of four bilayers of Ag (38 nm) and $SiO_2$ (179 nm). The simulated energy (*E*) versus in-plane momentum vector (*k*) mapping for (**b**) Transmission and (**d**) Reflection geometries. The circled points in Figures (**b**) and (**d**) represent the experimental data obtained from angle tuning reflection and transmission spectrum. (**c**) Simulated transmission and reflection spectrum of the photonic structure.

## 2.2. Nonlinear Optical Response in Metal-Dielectric Photonic Structures

The optical properties of the metal-dielectric structures are greatly influenced by the free valence electrons of the constituent metal. The nonlinear optical properties are mainly described by the ultrafast and very high nonlinear responses of the metal in a picosecond time scale. [4,8] Absorption plays a crucial role in enhancing the nonlinear optical response of MD photonic architecture. [4,7,28,29] Usually, the Drude model is used to relate the electron temperature to the dielectric

_____

function and hence the index of refraction. The ultrashort laser pulse heats the structure locally, which leads to the temperature-dependent Drude coefficients as follows. [4, 10]

$$\varepsilon(\omega) = \varepsilon_\infty - \frac{\omega_p(T)^2}{i\omega\gamma(T)}, \qquad (1)$$

where $\varepsilon_\infty$ is the sum of the interband contributions, $\omega_p$ is the bulk plasma frequency of metal, $\gamma$ is a frequency-independent damping parameter. The temperature dependence can be well interpreted using a two-temperature (2T) model for the metallic system. The main focus here is to investigate the nonlinear optical responses, which can effectively be characterized by the femtosecond time-resolved optical pump-probe spectroscopy. Further, a two-temperature model is incorporated into the transfer matrix method simulation to model the experimental findings qualitatively.

### 2.2.1. *Nonlinear Absorption Dynamics of Photonic Minibands: experiment and simulations*

The temporal evolution of the nonlinear optical response of metal-dielectric structure was investigated via femtosecond transient absorption spectroscopy using the optical pump-probe technique. Figure 2a shows the experimentally obtained transient difference absorption (*ΔA*) spectra at different time delays, pumped at 3.54 eV with pulse energy 0.2 μJ and probed by a weak-broadband white-light continuum spanned between 1.6 eV to 2.1 eV, covering the red-end photonic minibands. The experimental delay time versus probe energy transient absorption (ΔA) map is shown in Figure 2b, which captures the signature of the difference absorption (ΔA) of photonic minibands upon pump excitation The observed transient absorption responses of photonic minibands are the direct consequence of the photon density fluctuations as introduced by the coupled Fabry-Perot cavities. The redshift in the transient absorption (ΔA) peak position can be attributed to the optical phonon vibrations initiated by 3.54 eV pump excitation at the longer time scales.

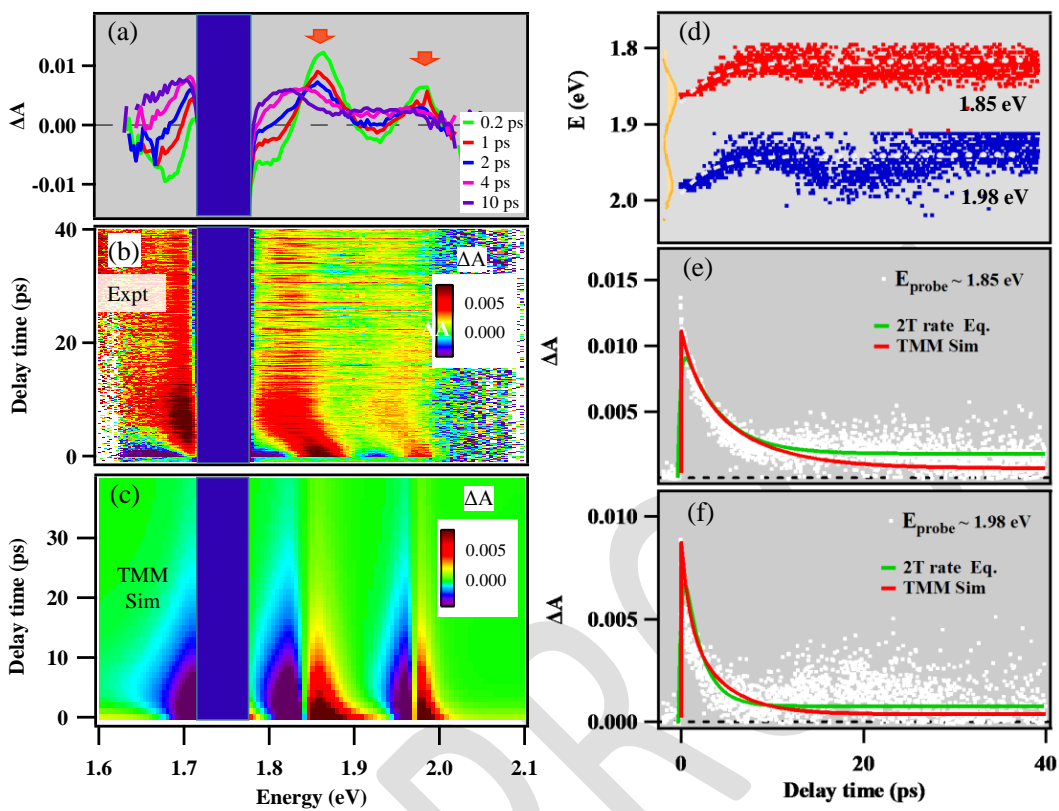

**Figure 2. (a)** Difference transient absorption (ΔA) spectra of (Ag/SiO$_2$)$_4$ metal-dielectric (MD) structure at various delay times, pump at 3.54 eV (0.2 µJ, 120 fs, 1 kHz) with broadband probe (1.6 to 2.1eV). **(b)** Experimental ΔA map of probe delay versus probe energy at 0.2 µJ pump energy. **(c)** Numerical simulations of ΔA map using Equation (2) based modified Transfer Matrix Method (TMM). **(d)** The ΔA peak shift of photonic minibands (1.85 and 1.98eV) as a function of probe delay (white dashed lines indicate the trend). Experimental ΔA dynamics (white dotted points) for probe energies at **(e)** 1.85 eV and **(f)** 1.98 eV along with the TMM based simulation (Equation (2)) and two-temperature (2T) rate equation (Equation (3)) based fits (solid red and green lines respectively). The dark blue shading in Figures 2(a-c) covers the harmonic of the pump signal.

In our previous studies, we have demonstrated that the main origin of nonlinearity as induced by high intense laser interaction is from electronic contribution. [7] In this MD structure, the third-order nonlinearity of SiO$_2$ is negligible thus the optical nonlinearity comes solely from the Ag intrinsic nonlinearity, which is enhanced manifold by the coupled cavity resonators. While modeling the transient absorption dynamics, the nonlinear response is attributed to the transient change of Ag dielectric function where the time evolution of the change in the optical constants,

_____

*viz.*, refractive index (*Δn*) and the extinction coefficients (*Δk*) of Ag, are incorporated into the TMM formalism.[4] The temporal evolution of *n* and *k* can be introduced by the following phenomenological expressions

$$\Delta n = \sum_{i=1}^{2} N_i e^{-t/\tau_i} \quad \text{and} \quad \Delta k = \sum_{j=1}^{2} K_j e^{-t/\tau_j}. \tag{2}$$

Where $\tau_i$ and $\tau_j$ are the decay constants derived from the experiment (Figures 2e, f), and $N_i$ and $K_j$ are adjustable constants. Figure 2c depicts the numerical simulations utilizing Equation (2) based transfer matrix method. By comparing Figures 2b and 2c, the simulated transient absorption (ΔA) of delay time versus probe energy mapping mimics and in good qualitative agreement with the experimental results. The temporal fittings of the experimental data (white dots in Figures 2e,f) are represented by solid red lines (Figures 2e,f) for photonic miniband maxima energies at 1.85 eV and 1.98 eV.

As mentioned, the ultrafast responses of these MD structures are dominated by coupled-cavity influenced Ag layers and undergo several dynamic processes involving electron-electron and electron-photon interactions. By largely considering these two interactions, a simplified two-temperature (2T) rate equation is implemented to analyze the temporal absorption dynamics evolution as follows. [4,5,30,31]

$$\Delta A = (\Delta A)_C + \left[ (\Delta A)_N e^{-(\alpha_{e-e}+\beta_{e-ph})t} + (\Delta A)_T (1-e^{-(\alpha_{e-e})t})e^{-(\beta_{e-ph})t} + (\Delta A)_L (1-e^{-(\beta_{e-ph})t}) \right], \tag{3}$$

where N, T, L and C stand for non-thermal (*N*) distribution, thermal distribution (*T*) of the electrons, lattice heating (*L*), and a long-lived offset (*C*) contributing factors. The simplified 2T model Equation (3) depends upon both electron-electron ($\alpha_{e-e}$) and electron-phonon ($\beta_{e-ph}$) coupling rates. Figures 2e and 2f contain the ΔA numerical fits of 2T model (solid red lines) dynamics, along with TMM numerical simulations, at energies 1.85 eV and 1.98 eV, respectively. The values of $\alpha_{e-e}$ and $\beta_{e-ph}$ turn out to be 3.5 ps$^{-1}$ and 0.25 ps$^{-1}$, respectively, when probed at 1.85 eV whereas, when probed at 1.98 eV the values are 3.1 ps$^{-1}$ and 0.5 ps$^{-1}$, respectively.

The energy relaxation process in metallic systems involves different time scales for various elementary processes such as *e-e* and *e-ph* interactions, which eventually decide the effective nonlinear dynamic responses of a composite metallic system.[6] Figure 3a displays a schematic representation of electron dynamics at different time scales involved upon ultrafast laser pulse excitation. The electron distribution is hugely modified upon femtosecond laser excitation that can be visualized into two separate components as Fermi (thermalized) and non-Fermi (non-thermalized) distributions.[31] The non-thermal electron distribution eventually thermalizes through the electron-electron (*e-e*) scattering process within 0.05 to 0.5 ps. The hot electron gas thermalizes with the surrounding lattice through the electron-phonon (*e-ph*) scattering process within 0.5 to 5 ps. Later, the system relaxes through a long-lived lattice cooling process on a time scale of 10 to 1000 ps. [4,8,9] Many complex metallic systems are analyzed through the two-temperature (2T) based ultrafast dynamics. Bigot et al. [6] reported the electron dynamics in copper (Cu) and silver (Ag) nanoparticles embedded in a transparent matrix utilizing the 2T model. Sun et al. [31] reported that the nonthermal electronic response in gold (Au) is found to be dominant features in electron thermalization dynamics (~0.5 ps) in the regime far from the interband transition threshold and close to the interband transition, the thermal electrons are found to be the contributing factor at a time scale of 1-2 ps.[31] Petros Farah et al. [4] reported the *e-e* scattering rate of 2.75 ps$^{-1}$ and *e-ph* scattering rate ranging from 1.5 to 0.5 ps$^{-1}$ in the plasmonic region of Au-polymer (PDMS) structures.[4] In the present case of (Ag/SiO$_2$)$_4$ MD system, the extracted values of $α_{e-e}$ and $β_{e-ph}$ are in well agreement with the previously discussed time scales of electron-electron and electron-phonon interactions. The optical field deposits differently at different penetration depths and excitation energies (Figure S2), thus the ultrafast dynamics and nonlinear optical responses of noble metals also vary accordingly. Therefore, the measured values of $α_{e-e}$ and $β_{e-ph}$ can be considered as the average value contributed by all the metal layers spaced by the neutral dielectric layer. It is also further noticed that this heat energy further broadens the electronic energy bands during the process of lattice cooling, which results in a red-shift in the transient absorption peak positions, [32] which is already evident in Figure 2d.

Pump energy-dependent transient absorption studies were also performed to observe the variation in $α_{e-e}$ and $β_{e-ph}$. Figure 3c shows the experimental ΔA temporal dynamics, and the 2T rate equation fits for the probe at 1.85 eV for varying pump energies ranging from 0.05 to 0.20 μJ. The electron-electron coupling rate ($α_{e-e}$) remains almost constant for different pump energies, whereas the electron-phonon coupling rate ($β_{e-ph}$) decreases with increasing pump energies (Figure 3d) as expected from the two-temperature model.[4,33] The thermal contribution, supposed to appear at a much longer time scale (> 1000 ps), is not considered here.

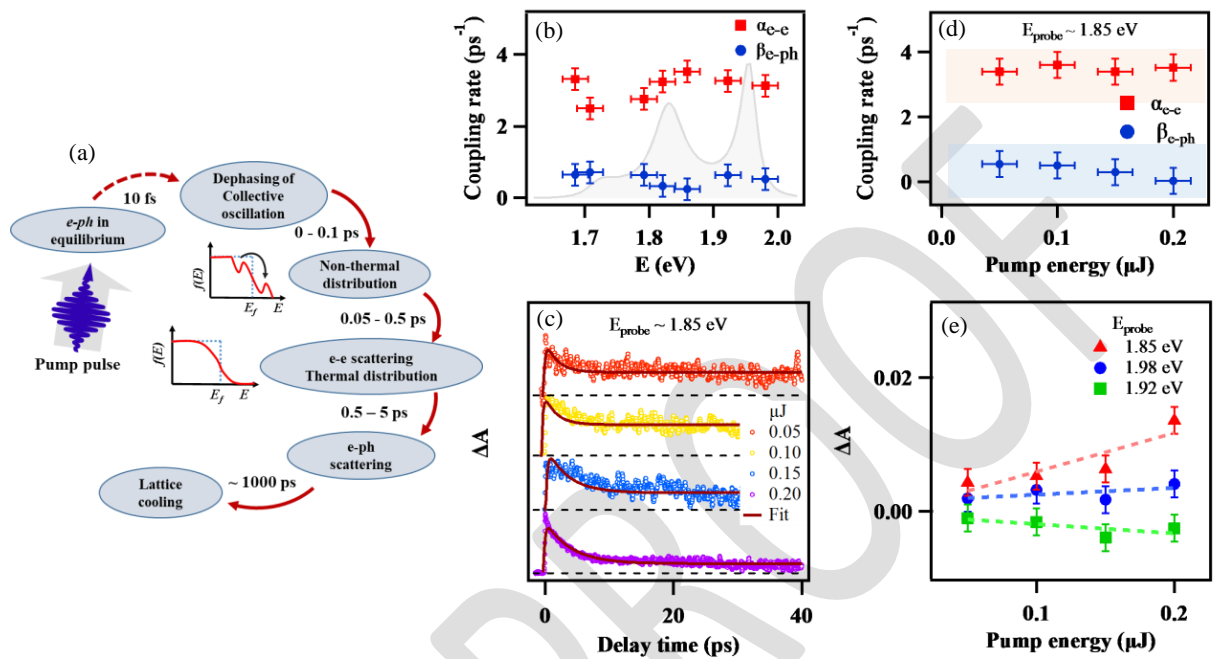

**Figure 3. (a)** Schematic representation of the temporal evolution of electron dynamics in the metal-dielectric (MD) system after pump absorption. **(b)** Plots of electron-electron ($\alpha_{e\text{-}e}$) and electron-phonon ($\beta_{e\text{-}ph}$) coupling rates of $(Ag/SiO_2)_4$ MD structure extracted from Figure 2(b) using 2T rate Equation (3) for different probe energies. The shaded region is the transmission spectra of the MD structure. **(c)** Pump energy dependent transient absorption ($\Delta A$) temporal dynamics at 1.85 eV along with 2T rate model fits and **(d)** extracted coupling rates ($\alpha_{e\text{-}e}$ and $\beta_{e\text{-}ph}$) from 2T rate Equation (3). **(e)** Plots of pump energy dependent $\Delta A$ peak amplitudes probed at 1.98 eV, 1.92 eV, and 1.85 eV. Dashed lines are guide to the eye.

The electron-phonon coupling ($\beta_{e\text{-}ph}$) rates strongly depend on the differences in the heat capacities between the electronic systems and the lattice. The increase in the pump energy leads to an increase in electron temperature, which results in the Fermi smearing effect and increases the electron heat capacities. Thus, the energy exchange between the electrons and the lattice decreases, which causes a decrease in $\beta_{e\text{-}ph}$ with increasing pump energies, as evident in Figure 3d. [4,33] For comparison, the transient absorption dynamics of the $(Ag/SiO_2)_1$ single bilayer and bare Ag thin film are also studied, and the results are depicted in the Supporting Information Figure S3.

## 2.2.2. Third-order Nonlinear Optical Measurements using Ultrafast Single-Beam Propagation Studies

To further support the ultrafast nonlinear optical responses of $(Ag/SiO_2)_4$ MD structure, the nonlinear responses at 3.54 eV pump energy were performed using a single Gaussian beam Z-scan method [7, 34]. For a focused Gaussian $TEM_{00}$ laser pulse propagating along the Z-direction, the normalized transmittance for open aperture (OA) and closed aperture (CA) Z-scan traces can be expressed as follows [7,34]

$$T_{OA}(z) = 1 - \frac{1}{2^{3/2}} \frac{\alpha_{2eff} I L_{eff}}{(1+(z/z_r)^2)}, \tag{4a}$$

$$T_{CA}(z) = 1 - \frac{4\Delta\phi(z/z_r)}{(1+(z/z_r)^2)(9+(z/z_r)^2)}, \tag{4b}$$

where z is the propagation distance, $z_r$ is called the Rayleigh length, $\alpha_{2eff}$ (m W$^{-1}$) represents the effective two-photon absorption (2PA) coefficients, $L_{eff}$ is the effective sample length that can be expressed as $L_{eff} = \frac{1-\exp(-\alpha_0 L)}{\alpha_0}$ (where $\alpha_0$ is the linear absorption coefficient and $L$ is the sample length). The nonlinear phase shift ($\Delta\phi$) is extracted from the closed aperture fit (Equation (4b)) to calculate the nonlinear refractive index ($n_2$) as $n_2 = \frac{|\Delta\phi|\lambda}{2\pi I L_{eff}}$ (m$^2$ W$^{-1}$) where λ represents the excitation wavelength of the laser pulse. The normalized OA Z-scan transmittance curves for $(Ag/SiO_2)_4$ MD structure are shown in Figure 4a at 3.54 eV (120 fs, 1 kHz) for various pulse energies, ranging from 0.02 to 0.35 μJ. The relative transmission at the focal point shows a reduction of transmittance with an increase of input beam intensity, suggesting reverse saturable absorption (RSA). The theoretical fits from Equation (4a) of the OA scans are shown in Figure 4a, along with the experimental data. The extracted nonlinear absorption coefficient ($\alpha_{2eff}$) monotonically decreases from 56 x 10$^{-7}$ to 2x10$^{-7}$ m W$^{-1}$ with the increase in the pump energies from 0.02 to 0.35 μJ (Figure 4b). The pump energy dependent trend suggests that the system undergoes different progressive excited-state absorption (ESA) mechanisms and subsequently

_____

relaxes through various scattering mechanisms, as discussed in Figure 3. The optical limiting behavior of (Ag/SiO$_2$)$_4$ MD structure is also shown in Figure 4c, which supports the potential applications for high power laser operations. These MD structures show a positive refractive nonlinearity (self-focusing effects), which is evident from the CA Z-scan transmission (Supporting Information Figure S4). The estimated nonlinear refractive index (n$_2$) value is about ~ (+)1.0x10$^{-19}$ m$^2$ W$^{-1}$ at pulse energy 0.35 µJ. Furthermore, the positive nonlinear refraction indicates that the electronic response will be in the picosecond time scale, [4,8] which further supports the transient absorption dynamics (Figure 3).

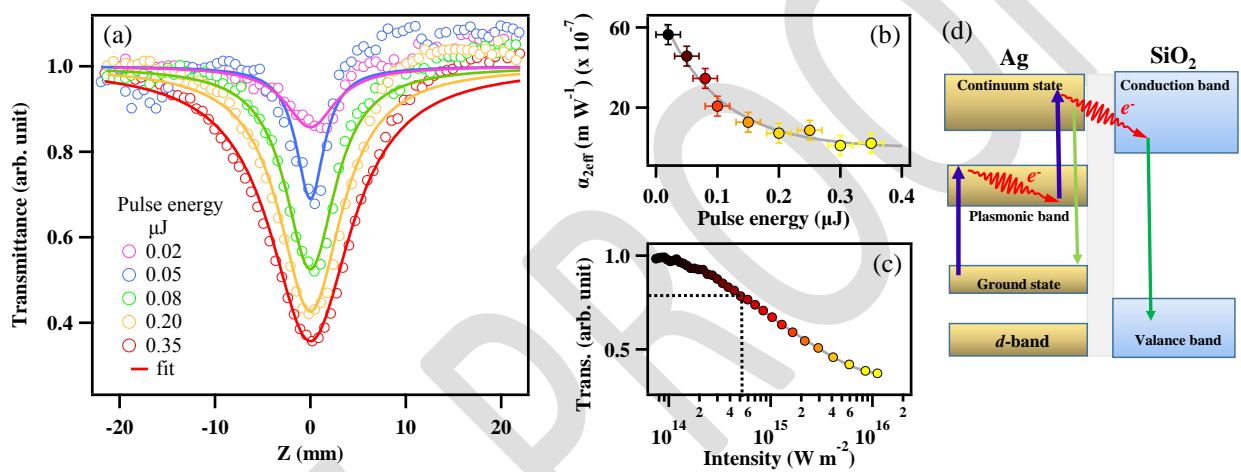

**Figure 4. (a)** Open aperture (OA) Z-scans of (Ag/SiO$_2$)$_4$ at 3.54 eV (120fs, 1KHz) for different pulse energies. Solid lines are the theoretical fit of Eq 4a. **(b)** The pulse energy variation of extracted effective nonlinear absorption coefficient ($\alpha_{2eff}$). The dashed line represents the trends. **(c)** Nonlinear optical limiting behavior at 0.35 µJ pulse energy. **(d)** Schematic representation of different electron excitation and relaxations in (Ag/SiO$_2$)$_4$ system upon 3.54 eV excitation.

The broad origin of nonlinear optical behavior in this metal-dielectric structure can be realized by invoking the energy levels of the coupled Ag/SiO$_2$ system. The conduction band electrons of Ag (free carriers) get promoted to the Ag plasmonic band (i.e., at ~ 3.1 eV) upon excitation with the photon of energy 3.54 eV. Then the intraband relaxation mechanism brings down the excited electrons to the bottom of the plasmonic band, and eventually, these plasmonic band electrons are further nonlinearly excited to the continuum of Ag with the support of pump photons of 3.54 eV. The first phenomenon is called the free-carrier absorption (FCA), and the subsequent one is the excited-state absorption (ESA). The SiO$_2$ matrix plays an important role in the relaxation of the excited electrons. The schematic representation of the energy levels of the system is depicted in Figure 4d. Thus, the RSA based nonlinear absorption mechanisms involved in the OA traces are mainly from free-carrier absorption (FCA) and various higher-order excited state absorptions (ESA). The femtosecond laser pulse interaction induced nonlinear optical responses of $(Ag/SiO_2)_4$ MD structures in the non-resonant region (~1.71 eV) are enhanced manifold due to the formation of coupled Fabry-Perot cavity resonators.[7] The nonlinear absorption coefficients reported at 1.71 eV are also provided in table 1.

For comparison, the single-beam Z-scan studies for $(Ag/SiO_2)_1$ single bilayer and bare Ag thin film have also been conducted at 3.54 eV, and the data are shown in the Supporting Information Figure S5. The extracted nonlinear absorption coefficients for $(Ag/SiO_2)_4$ are listed in table 1, along with $(Ag/SiO_2)_1$ and bare Ag thin film. The extracted third-order nonlinear absorption coefficients ($\alpha_{2eff}$) of $(Ag/SiO_2)_1$ structure and bare Ag thin film at 3.54 eV are about two orders of magnitude lower than that of the $(Ag/SiO_2)_4$ MD structure. Since the third-order nonlinear response of used silica ($\chi^{(3)}_{SiO_2}$) is much smaller than the silver ($\chi^{(3)}_{Ag}$), the enhanced features of the optical response can be attributed to the enhancement of Ag intrinsic nonlinearity by the coupled Fabry-Perot resonators.[7] Lepeshkin et al. [14] attributed the origin of the optical nonlinearity in noble metals to the Fermi smearing effect, which generally describes the modifications of the Fermi distribution of the electron gas upon excitation of the laser pulses. This sub-picosecond Fermi Smearing process is faster than the other thermal processes. In our previous communication, we have demonstrated manifold enhanced optical nonlinear absorption in $(Ag/SiO_2)_4$ MD structure in the non-resonant (NIR, 1.7 to 2.0 eV) spectral region, as compared to $(Ag/SiO_2)_1$ and bulk Ag thin films.[7] Suresh et al. also reported nonlinear optical features in the strongly absorbing region (at 2.48 eV) of $(Ag (16\ nm)/SiO_2(65\ nm))_5$ type metamaterial with nonlinear absorption coefficient of $(-)1.5 \times 10^{-5}$ m W$^{-1}$. [15]

**Table 1.** Third-order nonlinear absorption coefficients ($\alpha_{2eff}$) from the open aperture Z-scan traces of $(Ag/SiO_2)_4$, $(Ag/SiO_2)_1$, and bare Ag thin film at 3.54 eV. The nonlinear coefficients at probe energy 1.71 eV are also provided here, from Ref 7.

| Pulse energy (μJ) | Nonlinear absorption coefficient ($\alpha_{2eff}$) (m W$^{-1}$) (x 10$^{-7}$) $(Ag/SiO_2)_4$ | Nonlinear absorption coefficient ($\alpha_{2eff}$) (m W$^{-1}$) (x 10$^{-9}$) $(Ag/SiO_2)_1$ | Nonlinear absorption coefficient ($\alpha_{2eff}$) (m W$^{-1}$) (x 10$^{-9}$) *Bare Ag thin film* |
| --- | --- | --- | --- |

| At E = 3.54 eV | | | |
|---|---|---|---|
| 0.02 | 56.44 | 24.2 | 5.37 |
| 0.05 | 45.65 | 27.6 | 4.87 |
| 0.08 | 34.57 | 28.4 | 3.06 |
| 0.1 | 20.74 | 27.8 | 3.1 |
| 0.15 | 12.69 | 29.9 | 3.07 |
| 0.2 | 7.145 | --- | --- |
| 0.25 | 8.59 | --- | --- |
| 0.3 | 1.02 | --- | --- |
| 0.35 | 2.1 | --- | --- |
| At E = 1.71 eV [7] | | | |
| --- | $\alpha_{2eff} = 7.61 \times 10^{-9}$ m W$^{-1}$, $\alpha_{3eff} = 5.13 \times 10^{-23}$ m$^3$ W$^{-2}$ $\alpha_{4eff} = 1.15 \times 10^{-37}$ m$^5$ W$^{-3}$ at 20.00 GW cm$^{-2}$ | $\alpha_{2eff} = 2.42 \times 10^{-9}$ m W$^{-1}$ at 9.48 GW cm$^{-2}$ | $\alpha_{2eff} = 4.40 \times 10^{-9}$ m W$^{-1}$ at 0.10 GW cm$^{-2}$ |

## 2.3. Ultrafast Pulse Propagation through Metal-Dielectric Multilayers

Extremely high optical nonlinearity attracts for the investigation of the ultra-short pulse dynamics in metal-dielectric (MD) structures both numerically as well as experimentally. The soliton/soliton-plasmon formation studies have been performed in (metal/Kerr nonlinear medium/dielectric)$_1$ 1D structure [16] and cylindrical MD geometries [17] for transverse incidence and discrete solitons in longitudinal incidence.[35] In what follows, we present a simplistic numerical model to simulate the dynamics of ultra-short pulse in such MD structure.

The MD structure under investigation shows a giant optical nonlinearity at the photonic minibands.[7] The ultrafast and very high nonlinear response of the metallic system significantly modifies the propagation of the pump pulse through the MD structure. The dynamics of ultrafast pulse propagation is characterized by the wave equation, which is given by [36]

$$\nabla^2 E - \frac{1}{c^2}\frac{\partial^2 E}{\partial t^2} = \mu_0 \left( \frac{\partial^2 P_L}{\partial t^2} + \frac{\partial^2 P_{NL}}{\partial t^2} \right), \tag{5}$$

where $P_L$ and $P_{NL}$ are the induced linear and nonlinear polarization, respectively, $E(r,t)$ is the electric field, $c$ is the velocity of light in free space, and $\mu_0$ represents the free space permeability.

Under the slowly varying envelope approximation and paraxial propagation conditions, the pulse propagation in the one-dimensional photonic crystal can be described by a simplified version of modified scalar nonlinear Schrödinger equation (NLSE), which is given by [22,23,37-40]

$$\frac{\partial E}{\partial z} = -i\frac{\beta_2}{2}\frac{\partial^2 E}{\partial t^2} - \frac{\alpha}{2}E - i\gamma_{eff}|E|^2 E - a_0\frac{\alpha_{2eff}}{2}|E|^2 E - b_0\frac{\alpha_{3eff}}{2}|E|^4 E - c_0\frac{\alpha_{4eff}}{2}|E|^6 E, \quad (6)$$

where envelope $E(z,t)$ of the pulse is a function of both propagation distance ($z$) and time ($t$). The $\gamma_{eff} = \frac{n_2 \omega_0}{c A_{eff}}$ is effective nonlinear refractive index coefficient and is related to Kerr nonlinearity (nonlinear refractive index, $n_2$). $\omega_0$ is the angular frequency, $c$ is the velocity of light in free space, and $A_{eff}$ denotes the effective modal area. [36] $\beta_2$ is called group velocity dispersion that is defined as $\beta_2 = \frac{d}{d\omega}(\frac{1}{v_g})$ and is responsible for the pulse broadening in time, and $\alpha$ is the linear absorption coefficient. $\alpha_{2eff}$, $\alpha_{3eff}$, and $\alpha_{4eff}$ are two-, three- and four-photon absorption coefficients, respectively. The $a_0$, $b_0$, and $c_0$ are multiplication coefficients to their respective terms. Here, we assume the values of all these coefficients to be unity. To delineate the roles of the individual terms with these multiplicative coefficients, the values of either of these coefficients or multi-photon absorption coefficients may be varied. The nonlinear Schrödinger equation of Equation (6) is solved numerically using the Split step Fourier transform method. [36]

The tunable input Gaussian laser pulse of 120 fs and repetition rate of 1 kHz is launched perpendicularly to the plane of the one-dimensional metal-dielectric stack. The numerical investigation was performed at two energies, 1.71 eV and 3.54 eV (indicated in Figures 5 and 6). Figures 5 and Figure 6 depict the pulse evolution dynamics along the length (Z- direction) of the photonic structure at 3.54 eV as well as 1.71 eV for input pulse peak power of $17 \times 10^5$ W. The pulse propagation at 3.54 eV is modelled incorporating the nonlinear coefficients ($\alpha_{2eff} = 5.64 \times 10^{-6}$ m W$^{-1}$, $n_2 = 1.0 \times 10^{-19}$ m$^2$ W$^{-1}$), which are taken from the experimental values from Z-scan measurements (see. Figure 4 and Supporting Information Figure S4). On the other hand, the pulse propagation at 1.71 eV is modelled by incorporating the multi-photon absorption terms in Equations (5). Experimentally derived values of different simulation parameters at 1.71 eV pulse are taken from our previous studies [7] the linear absorption loss coefficient $\alpha = 24.8 \times 10^5$ m$^{-1}$, two-, three- and four-photon absorption coefficients are respectively $\alpha_{2eff} = 7.61 \times 10^{-9}$ m W$^{-1}$, $\alpha_{3eff} = 5.13 \times 10^{-23}$ m$^3$ W$^{-2}$ $\alpha_{4eff} = 1.15 \times 10^{-37}$ m$^5$ W$^{-3}$ and Kerr nonlinear refractive index ($n_2$) is $1.26 \times 10^{-15}$ m$^2$ W$^{-1}$. For both these energies, the total propagation distance ($Z$) is $1.1 \times 10^{-6}$ m, which is the width of the real multilayer structure.

Figure 5a shows 3D maps of intense ultrafast Gaussian pulse at 3.54 eV along the length ($z$) of the MD structure. While Figure 5b represents respective 2D spectral representations. Further, simulations were also performed for different nonlinear absorption and different Kerr coefficients. In the first set of simulations, we varied the nonlinear absorption loss ($\alpha_{2eff}$), and other terms, including the nonlinear Kerr term ($\gamma_{eff}$), were kept constant. In the second set of simulations, the nonlinear Kerr term was varied, and we kept the other terms unchanged. Figures 5c shows the output pulse spectrum for varying nonlinear absorption loss ($\alpha_{2eff}$) for a fixed $\gamma_{eff}$ (= 0.0086 m$^{-1}$ W$^{-1}$) at 3.54 eV pulses. On the other hand, Figure 5d represents the spectral pulse profiles for varying Kerr nonlinearity ($\gamma_{eff}$) for a fixed $\alpha_{2eff}$ (= 5.64x10$^{-6}$ m W$^{-1}$) for pulses centered at 3.54 eV.

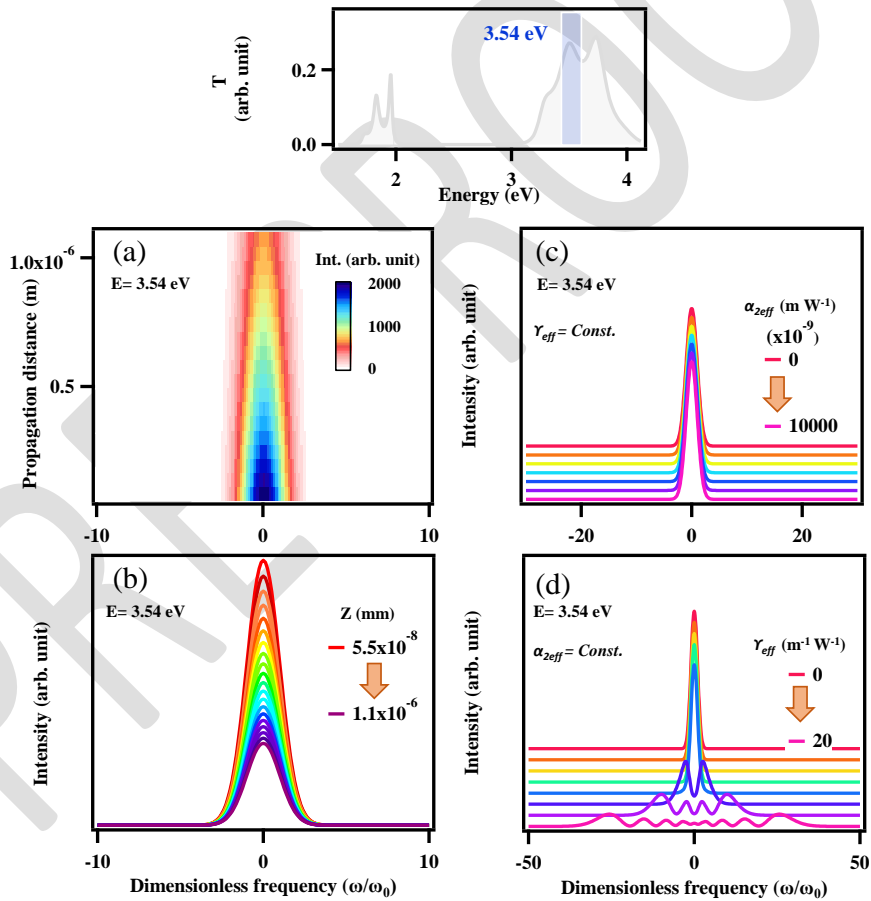

**Figure 5.** Ultrafast nonlinear pulse propagation in metal-dielectric (MD) photonic structure along the length of the structure at resonant energy 3.54 eV. **(a)** Pulse intensity map along the propagation length ($z$) and (b) pulse spectra at different pulse propagation length. Linear absorption coefficient $\alpha$ = 11.2x10$^5$ m$^{-1}$, two-photon absorption coefficients, $\alpha_{2eff}$ = 5.64x10$^{-6}$ m W$^{-1}$, effective nonlinear refractive index coefficient, $\gamma_{eff}$ = 0.0086 m$^{-1}$ W$^{-1}$. (c) Pulse profiles at different $\alpha_{2eff}$ values for a

_____

fixed $\gamma_{eff}$ (= 0.0086 m$^{-1}$ W$^{-1}$). (d) Pulse profiles at different $\gamma_{eff}$ values for a fixed $\alpha_{2eff}$ (= 5.64x10$^{-6}$ m W$^{-1}$). The propagation length of the MD structure ($z$) is 1.1x10$^{-6}$ m, and pulse peak power is 17x10$^5$ W. The transmission spectrum shown at the top is an indicative.

In case of non-resonant excitation at 1.71 eV, a drastic change in the pulse propagation dynamics is observed. Figure 6a shows the 3D map of intense ultrafast Gaussian pulse at 1.71 eV along the length ($z$) of the MD structure. While Figure 6b represents respective 2D spectral representations. Figures 6c shows the output pulse spectrum for varying nonlinear absorption loss ($\alpha_{2eff}$) for a fixed value of $\alpha_{3eff}$ (= 5.13 × 10$^{-23}$ m$^3$ W$^{-2}$), $\alpha_{4eff}$ (=1.15 × 10$^{-37}$ m$^5$ W$^{-3}$), and $\gamma_{eff}$ (= 12.14 m$^{-1}$ W$^{-1}$) at 1.71 eV pulses. On the other hand, Figures 6d represents the spectral pulse profiles for varying Kerr nonlinearity $\gamma_{eff}$ for a fixed $\alpha_{2eff}$ (= 7.61x10$^{-9}$ m W$^{-1}$), $\alpha_{3eff}$ (= 5.13 × 10$^{-23}$ m$^3$ W$^{-2}$), $\alpha_{4eff}$ (=1.15 × 10$^{-37}$ m$^5$ W$^{-3}$) for a pulse centered at 1.71 eV. As the pulse propagates along the length of the structure, the pulse energy is attenuated due to the significant linear and nonlinear losses. This results in the reduction in pulse peak power with propagation, as observed in Figures 5a-6a. In addition, due to the relatively stronger nonlinearity, the significant spectral broadening is observed at 1.71 eV. A significant spectral broadening and new frequency generation are observed due to the variations of Kerr coefficients at both the energies, as depicted in Figures 5d and 6d. Whereas the variation of $\alpha_{2eff}$ does not affect the pulse propagation dynamics, as revealed in Figures 5c and 6c. Therefore, in the present case, the Kerr coefficient plays a significant role in spectral broadening and new frequency generation.

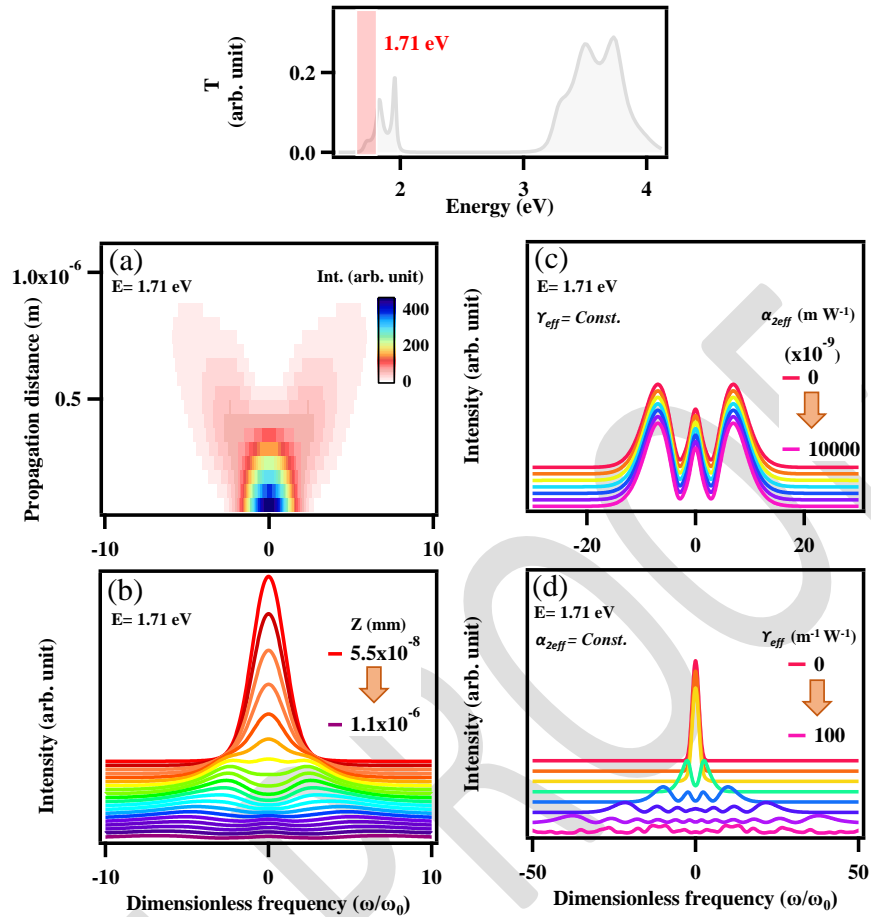

**Figure 6.** Ultrafast nonlinear pulse propagation in metal-dielectric (MD) photonic structures along the length of the structure at non-resonant energy 1.71 eV. **(a)** Pulse intensity map along the propagation length (*z*) and **(b)** pulse spectra at different pulse propagation length. The linear absorption coefficient α = 24.8x10$^5$ m$^{-1}$, two-photon $α_{2eff}$ = 7.61x10$^{-9}$ m W$^{-1}$, three-photon $α_{3eff}$ = 5.13 × 10$^{-23}$ m$^3$ W$^{-2}$ and four-photon absorption coefficient $α_{4eff}$ = 1.15 × 10$^{-37}$ m$^5$ W$^{-3}$. The effective nonlinear refractive index coefficient $γ_{eff}$ =12.14 m$^{-1}$ W$^{-1}$. **(c)** Pulse profile at different $α_{2eff}$ values for a fixed value of $α_{3eff}$, $α_{4eff}$, and $γ_{eff}$. (d) Pulse profile at different $γ_{eff}$ values for a fixed $α_{2eff}$, $α_{3eff}$ and $α_{4eff}$. The propagation length of the MD structure (*z*) is 1.1x10$^{-6}$ m and pulse peak power is 17x10$^5$ W. The transmission spectrum shown at the top is an indicative.

Owing to losses, intensity lowering with propagation is also observed in the time domain as depicted in the Supporting Information (Figure S6). As compared to pulse at 1.71 eV, the pulse at 3.54 eV does not reveal any significant variation in temporal as well as in the spectral domain except for the attenuation. This is obviously due to the smaller values of nonlinear parameters at 3.54 eV. Perhaps for 3.54 eV propagation, at higher peak powers (>10$^5$ W) the Kerr nonlinearity is expected to be higher along with the possibility of higher-order nonlinear terms. However, given the fact of practical purposes, such high powers are not been used here.

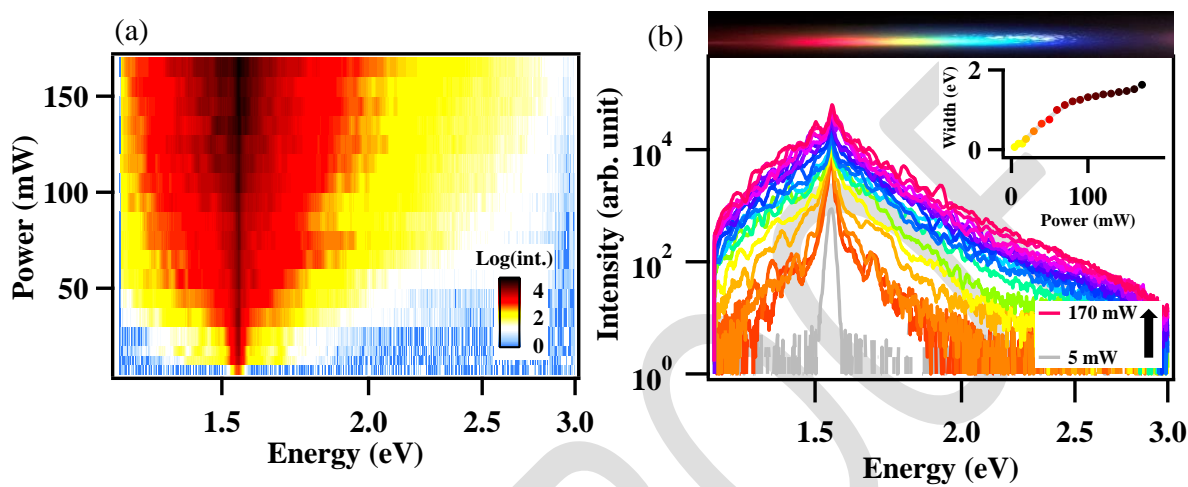

**Figure 7.** Femtosecond laser pulse (at 1.55 eV) broadening at various excitation pump powers (1 kHz, 120 fs). **(a)** Pump power versus wavelength intensity map and **(b)** corresponding spectra at various powers. The inset shows the variation of the pulse broadening (spectral width) with the input pump power.

To exemplify the pulse broadening effect, we performed an experiment where femtosecond laser pulse at 1.55 eV (800 nm) is used on $(Ag/SiO_2)_4$ MD structure. The laser beam is tightly focused onto the sample, and the propagated spectral profiles after MD structure are monitored with increase in the input laser average power. Figure 7a depicts the transmitted spectral intensity map of the generated spectrum at various pump powers. The pulse spectral broadening, spanning over > 600 nm with various laser power, is shown in Figure 7b. While further experiments are needed, the high nonlinearity is responsible for spectral broadening, which qualitatively mimics the simulation predictions mentioned previously.

## 3. Conclusion

In this paper, ultrafast laser pump-induced huge modulation of nonlinear optical responses modified by the metal Fermi electron distributions, are demonstrated in the one-dimensional metal-dielectric (MD) $(Ag/SiO_2)_4$ multilayer system. The ultrafast temporal evolution of the absorption dynamics has been investigated in the NIR photonic miniband spectral region, which are invoked

_____

by coupled-cavity resonator effects. Results are convincingly explained by the two-temperature (2T) rate equation and modified transfer matrix method analysis. The detailed energy relaxation dynamics of laser-heated electrons are described by incorporating the electron-electron (e-e) and electron-phonon (e-ph) interactions at picosecond time scales. The average values of electron-electron ($α_{e-e}$) and electron-phonon ($β_{e-ph}$) coupling rates are measured to be 3.1 ps$^{-1}$ and 0.53 ps$^{-1}$ respectively. The modified transfer matrix numerical simulations strongly correlate with the observed temporal behavior in $(Ag/SiO_2)_4$ MD structure. The nonlinear optical measurements further demonstrate that the intrinsic electronic nonlinearity of Ag is hugely enhanced by the coupled Fabry-Perot MD resonators. The enhanced nonlinear reverse saturation of absorption (RSA) at the pump energy (3.54 eV) is due to the free-carrier absorption (FCA) and excited-state absorption (ESA) processes. The effective nonlinear absorption of the $(Ag/SiO_2)_4$ MD structure is enhanced by about two orders of magnitude as compared to that of $(Ag/SiO_2)_1$ single bilayer and bare Ag thin film. Furthermore, direct implications of the nonlinear effect on the ultrafast pulse propagation dynamics were studied numerically by a pulse propagation model. The numerical experiments well established the feasibility of new frequency generations and impart a better understanding of nonlinear pulse propagation through these novel wavelength-ordered materials. Relatively higher optical Kerr nonlinearity makes these structures of special interest to observe nonlinear solitons. We envisage that such one-dimensional photonic structures can be used to tailor the nonlinear optical response for the manipulation of ultrafast pulse propagation dynamics and may find new applications in the high-intensity laser pulse-based nonlinear optical devices.

## 4. Materials Fabrications and Experimental Setup

*Sample Preparation*: The 1D $(Ag/SiO_2)_4$ structure was fabricated using alternative DC and RF sputtering techniques on BK7 glass substrate (Figure 1a). The detailed fabrication procedures can be found elsewhere. [7,41,42] The pre-calibrated deposition rates of Ag ( 38 nm) and SiO$_2$ (179 nm) are 0.20 nm s$^{-1}$ and 0.09 nm s$^{-1}$ in the presence of argon (Ar) gas at 1.2x10$^{-2}$ and 5.5 × 10$^{-3}$ millibar, respectively. Three different types of samples were fabricated (a) four bilayers $(Ag/SiO_2)_4$, (b) single bilayer $(Ag/SiO_2)_1$ and (c) bare Ag thin film.

_____

*Femtosecond Transient Absorption Spectroscopy:* The ultrafast transient absorption spectroscopy (pump-probe) was performed using the setup as shown in Figure S8 (Supporting Information). We used a 3.54 eV femtosecond pump pulse (120 fs, 1 kHz) from an optical parametric amplifier (OPA), which is fed by a fundamental 1.55 eV (120 fs, 1 kHz) femtosecond laser of a regenerative Ti:Sapphire amplifier. A weak broadband white-light continuum (1.2 to 3 eV) is used as a probe beam from a calcium fluoride ($CaF_2$) crystal, which was pumped by a part of fundamental 1.55 eV fs laser, which undergoes through a mechanical delay stage. The temporal and spatial overlapped are maintained on the sample surface between the pump and probe pulses to monitor the transient absorption signal. The pump-probe experiment was performed at ambient conditions.

*Femtosecond Z-scan measurements:* The nonlinear absorption and nonlinear refraction were investigated using the single-beam Z-scan technique at open aperture (OA) and closed aperture (CA) configurations using 3.54 eV Gaussian pulse of 120 fs pulse duration and 1 kHz repetition rate.[7] A convex lens of 150 mm focal length was used to focus the laser beam in a tight focus geometry. The silicon photodetectors, along with the motorized linear translational stage, have been synchronized for data acquisition. In OA Z-scan measurement, all the transmitted beam has been fully collected by a photodiode, whereas in CA measurement, an aperture is used before the photodiode. The experimental Z-scan traces were fitted using the least square fit method using equations mention in the main text. Further details of the experiment can be found in our previous communications.[7,43,44]

**Supporting Information**

The supplementary material entails a description of VU-Vis-NIR linear optical measurements and absorption dynamics of the $(Ag/SiO_2)_1$ single bilayer and bare Ag thin films, time-domain dynamics of pulses in the 1D stack from the numerical model, and the experimental set-up for the transient absorption spectroscopy.

**Acknowledgments**

The authors acknowledge DST-FIST funding for establishing the Ultrafast Optics facility (UFO). This work is partly supported by SERB-DST (Govt. of India) and Royal Society (UK) funding. The authors acknowledge Professor Anurag Sharma, Coordinator UFO, IIT Delhi, for useful suggestions. The authors also thank Prof. RB Gangineni of Pondicherry university for his help in the fabrication. The authors are thankful to Prof. Jeremy Baumberg, UK, for initial discussions and support. JNA acknowledges DST-INSPIRE for research fellowship.

**Conflict of Interest**

The authors declare no conflict of interest.

# Supporting Information

# Ultrafast nonlinear pulse propagation dynamics in metal-dielectric periodic photonic architectures

*Jitendra Nath Acharyya[1], Akhilesh Kumar Mishra[2*], D. Narayana Rao[3], Ajit Kumar[1], and G. Vijaya Prakash[1*]*

[1]*Nanophotonics Lab, Department of Physics, Indian Institute of Technology Delhi, New Delhi-110016, India*

[2]*Department of Physics, Indian Institute of Technology Roorkee, Roorkee-247667, India*

[3]*School of Physics, University of Hyderabad, Hyderabad-500046, India*

## 1. Linear optical responses of (Ag/SiO$_2$)$_1$ single bilayer and bare Ag thin film

The linear optical properties of (Ag/SiO$_2$)$_1$ single bilayer structure composed of silver (Ag, 38 nm) and fused silica (SiO$_2$, 179 nm) and bare Ag thin film (38 nm) were characterised using VU-Vis-NIR linear transmission and reflection measurements. The experimental results were correlated using transfer matrix method (TMM).

[*] Corresponding Author : akhilesh.mishra@ph.iitr.ac.in (AKM); prakash@physics.iitd.ac.in (GVP)

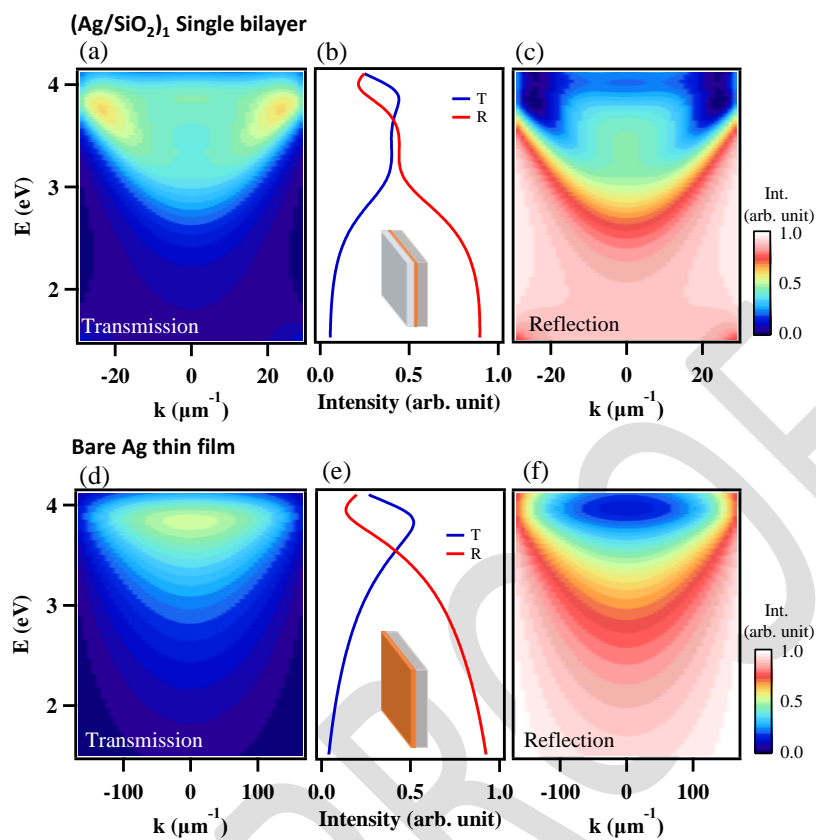

**Figure S1.** Simulated energy (*E*) versus in-plane momentum vector (*k*) mapping of $(Ag/SiO_2)_1$ single bilayer metal-dielectric (MD) structure for **(a)** Transmission and **(c)** Reflection geometries. **(b)** Simulated Transmission and Reflection Spectra of the $(Ag/SiO_2)_1$ single bilayer MD structure. Figures **(d)**, **(e)** and **(f)** are respective representations for bare Ag film of 38 nm thickness.

The detailed energy band structures of $(Ag/SiO_2)_1$ MD structure and Ag thin film are depicted in Figure S1. The reflection and transmission spectrum shows the signature of the strong interband absorption of Ag at about 3.85 eV.[1, 2] Unlike $(Ag/SiO_2)_4$ MD structure, $(Ag/SiO_2)_1$ and Ag exhibit a completely reflective spectrum in the visible and near-infrared regions.

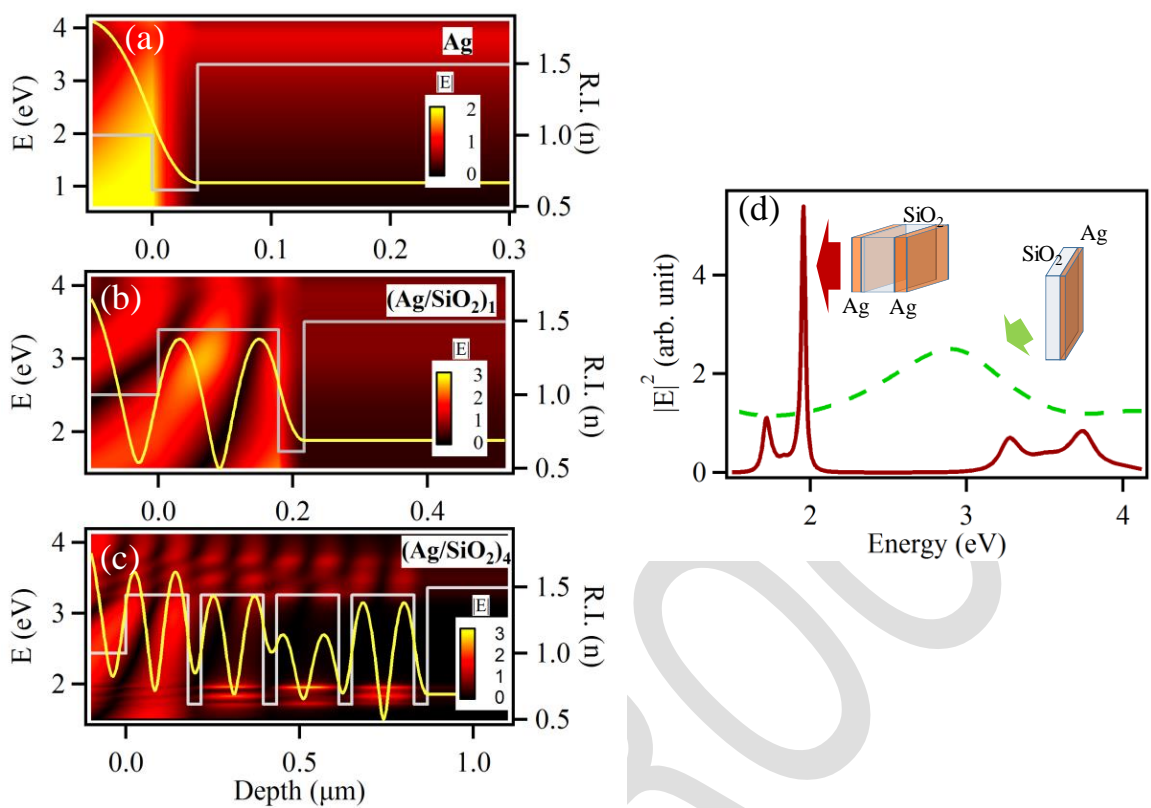

**Figure S2.** Spatial optical field ($|E|$) depth profile maps along with refractive index profile (white lines) and spatial optical field ($|E|$) profiles at 3.54 eV (yellow lines) for **(a)** bulk Ag thin film, **(b)** (Ag/SiO$_2$)$_1$ metal-dielectric structure, **(c)** (Ag/SiO$_2$)$_4$ metal-dielectric structure. **(e)** Optical field ($|E|^2$) deposited in the SiO$_2$ layer for (Ag-SiO$_2$-Ag) and (SiO$_2$-Ag) configurations.

The optical field deposition inside each layer of the metal-dielectric structure is studied using the spatial optical field ($|E|$) maps from TMM simulations (Figures S2a-S2c). Unlike (Ag/SiO$_2$)$_1$ and Ag thin film, a huge field confinement is observed in each SiO$_2$ layers of (Ag/SiO$_2$)$_4$ structure because of Ag-SiO$_2$-Ag configuration, which act as coupled *Fabry–Pérot* resonators that eventually confine the photons inside each SiO$_2$ layers. The spatial optical field ($|E|$) along the depth of the structure for the pump photon (3.54 eV) is depicted as yellow curves. in Figure S2a-2c. The optical field energy ($|E|^2$) depositions inside the SiO$_2$ layer are shown in Figure S2d for Ag-SiO$_2$-Ag and Ag-SiO$_2$ configurations.

## 2. Transient absorption dynamics of (Ag/SiO$_2$)$_1$ and bare Ag thin film

Transient absorption spectroscopy of (Ag/SiO$_2$)$_1$ MD structure and bare Ag thin film have been performed using 3.54 eV (120 fs, 1 kHz) as pump of 0.1 µJ pulse energy and broadband white-light continuum as probe. The observed transient absorption spectra show the signature of Ag plasmonic decay dynamics in both the photonic structures. Figure S3a shows the difference transient absorption (ΔA) spectra probed between 2 to 3.1 eV. The negative broad transient absorption (ΔA < 0) band centred at ~ 2.7 eV can be attributed to the Plasmon bleach dynamics. [3,4] The pump-probe delay time versus probe energy ΔA map of $(Ag/SiO_2)_1$ is depicted in Figure S3b. The temporal decay dynamics at 2.97 eV is depicted in Figure S3c. The extracted electron-electron coupling rate ($α_{e-e}$) and the electron-phonon coupling rate ($β_{e-ph}$) are 3.2 and 0.6 $ps^{-1}$, respectively for $(Ag/SiO_2)_1$ MD photonic structure. [5] Similar behaviour is observed for Ag thin film (Figures S3d,e). The transient absorption spectra of Ag thin film (Figure S3d) show a broad plasmon bleach signal. Figure S3f shows the temporal dynamics of the Plasmon beach signal at 2.85 eV. The extracted electron-electron coupling rate ($α_{e-e}$) and the electron-phonon coupling rate ($β_{e-ph}$) are 1.1 and 0.4 $ps^{-1}$, respectively for bare Ag thin film.

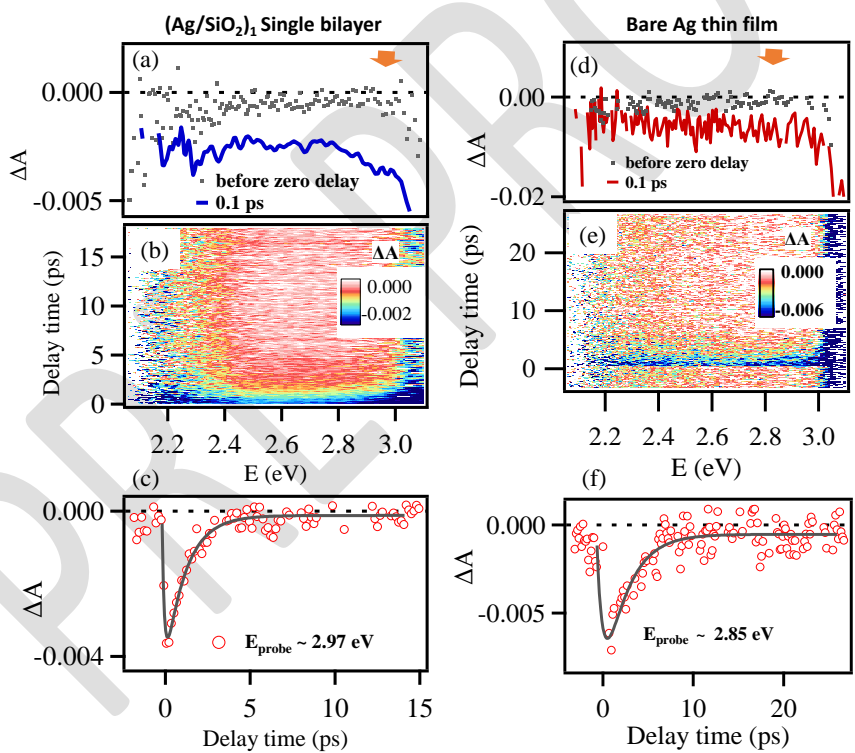

**Figure S3.** **(a)** Difference transient absorption (ΔA) spectra of $(Ag/SiO_2)_1$ metal-dielectric (MD) structure at 0.1 ps delay time, probing in the range of 2 to 3.1 eV for pumping at 3.54 eV (120 fs, 1 kHz) with 0.1 μJ pulse energy. **(b)** Experimental delay time versus probe energy ΔA map of $(Ag/SiO_2)_1$ MD structure pumping with 3.54 eV (0.1 μJ pulse energy). **(c)** Experimental ΔA temporal dynamics of $(Ag/SiO_2)_1$ MD structure probed at 2.97 eV (solid line represents the theoretical fit using two-temperature (2T) rate equation). **(d)** The ΔA spectra of bare Ag thin film at 0.1 ps delay time, probing in the range of 2 to 3.1 eV for pumping at 3.54 eV (120 fs, 1 kHz) of 0.1 µJ pulse energy. **(e)** Experimental delay time versus energy ΔA spectral map of bare Ag thin film pumping at 3.54 eV (0.1 µJ pulse energy). **(f)** The ΔA temporal dynamics of bare Ag thin film probed at 2.85 eV (solid line represents the theoretical fit using 2T rate equation).

### 3. Nonlinear refraction of (Ag/SiO$_2$)$_4$ MD structure at 3.54 eV

The details of open aperture Z-scan studies are given in the main text. Figure S4 depicts closed aperture (CA) Z-scan trace of (Ag/SiO$_2$)$_4$ MD photonic structure performed at 3.54 eV (120 fs, 1 kHz) laser pulse excitation of 0.35 µJ pulse energy. The CA Z-scan trace shows (Figure S4) a pre-focal dip with a post-focal peak which is the signature of positive nonlinear refraction (self-focusing behaviour) originating from the electronic Kerr nonlinearity. The solid line shows the theoretical fit using Equation (4b). The extracted nonlinear refractive index ($n_2$) is (+)1.0 x 10$^{-19}$ m$^2$ W$^{-1}$.

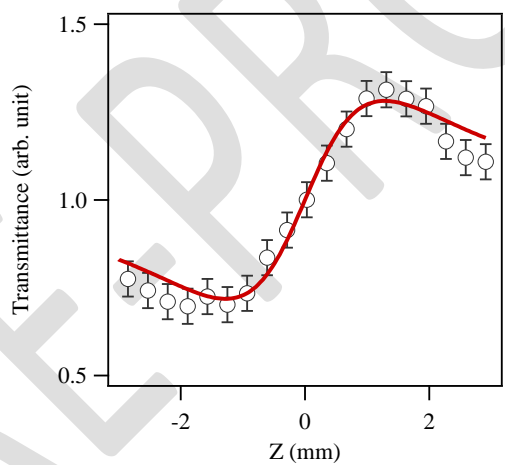

**Figure S4.** Closed aperture trace of (Ag/SiO$_2$)$_4$ with 3.54 eV (120 fs, 1 kHz) laser pulse of 0.35 µJ pulse energy (the solid red line represents the theoretical fit).

### 4. Nonlinear absorption of single bilayer (Ag/SiO$_2$)$_1$ and pure Ag thin film at 3.54 eV

Open aperture (OA) Z-scan traces of (Ag/SiO$_2$)$_1$ single bilayer and bare Ag thin film exhibit reverse saturable absorption (RSA) behaviour upon excitation at 3.54 eV (120 fs, 1 kHz) laser pulse. The experiments were performed with varying input laser pulse energies ranging from 0.02 to 0.15 µJ. The normalized OA traces of (Ag/SiO$_2$)$_1$ MD structure and pure Ag thin films are depicted in Figure S5a and Figure S5c. The solid lines represent the theoretical fit using Equation (4a). The variation in extracted nonlinear absorption coefficients ($\alpha_{2eff}$) with excitation pulse energies are shown in Figure S5b and S5d. The observed nonlinear absorption coefficients ($\alpha_{2eff}$) are in the order of ~ $10^{-9}$ m W$^{-1}$ in both the samples, which are two orders of magnitude lower than (Ag/SiO$_2$)$_4$ MD structure (~$10^{-7}$ m W$^{-1}$).

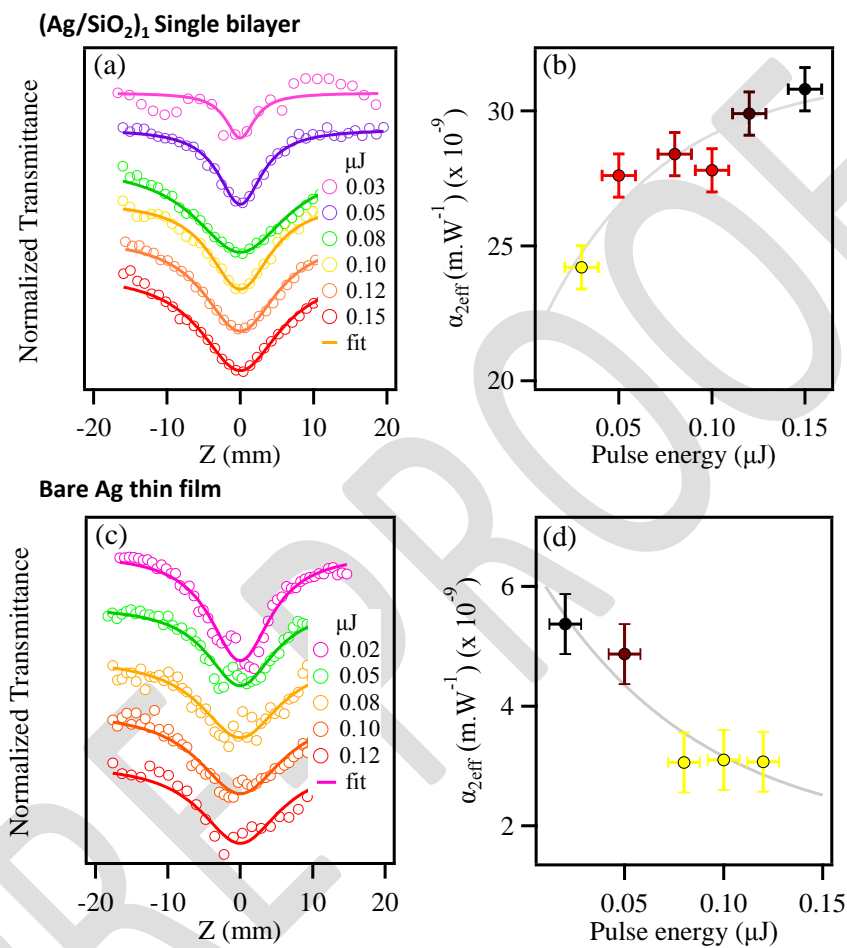

**Figure S5. (a)** Open aperture (OA) Z-scan traces of (Ag/SiO$_2$)$_1$ at 3.54 eV (120 fs, 1 kHz) laser pulse for various input pulse energies. The solid lines represent the theoretical fit using Equation (4a) (all the plots are shifted along the y-axis for clarity). **(b)** Plots of effective nonlinear absorption coefficient ($\alpha_{2eff}$) of (Ag/SiO$_2$)$_1$ structure extracted from the OA Z-scan traces from Figure S5a. The dashed line represents the trend. **(c)** The OA Z-scan traces of bare Ag thin film at 3.54 eV (120 fs, 1 kHz) excitation for varying input pulse energies (plots are shifted along the y-axis for clarity). The solid lines represent the theoretical fit for open aperture traces using Equation (4a). **(d)** Plots of $\alpha_{2eff}$ of bare Ag thin film extracted from the OA Z-scan traces from Figure S5c. The dashed line represents the trend.

## 5. Temporal dynamics of ultrafast pulse propagation through $(Ag/SiO_2)_4$ photonic structure

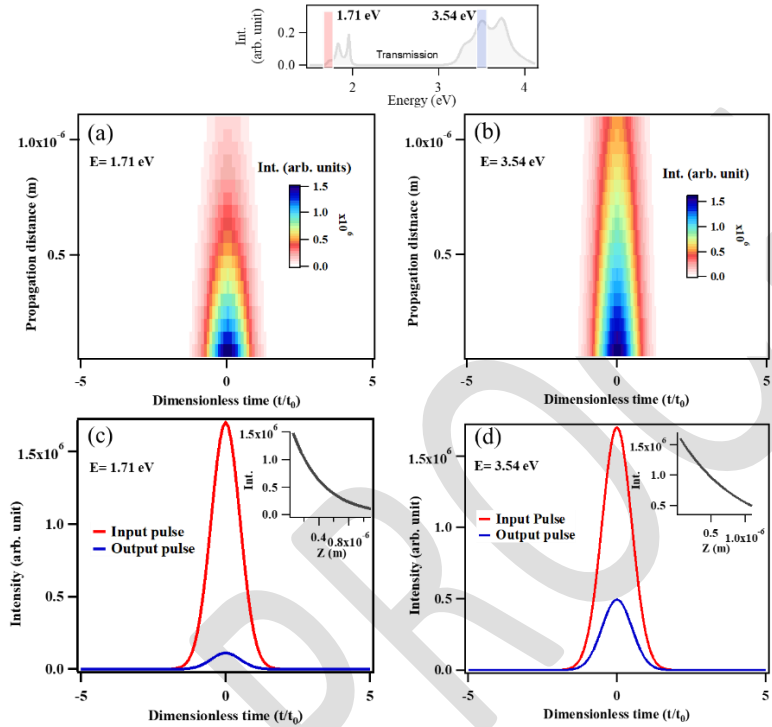

**Figure S6.** Ultrafast nonlinear pulse propagation in $(Ag/SiO_2)_4$ metal-dielectric (MD) photonic structures along the length of the structure. The temporal domain 3D intensity evolution of the pulse at **(a)** non-resonant energy 1.71 eV and **(b)** resonant energy 3.54 eV. Temporal domain input and output pulse profiles at **(c)** non-resonant energy 1.71 eV and **(d)** resonant energy 3.54 eV. The insets of (c)-(d) show the variation of the pulse temporal peak power along the length of the MD structure. The temporal axis is made dimensionless by the input FWHM width of the pulse (i.e., $t_0$ =120 fs). The nonlinear coefficients (experimentally derived) utilised in the simulation at 1.71 eV pulse propagation are $\alpha = 24.8 \times 10^5$ m$^{-1}$, $\alpha_{2eff} = 7.61 \times 10^{-9}$ m W$^{-1}$, $\alpha_{3eff} = 5.13 \times 10^{-23}$ m$^3$ W$^{-2}$, $\alpha_{4eff} = 1.15 \times 10^{-37}$ m$^5$ W$^{-3}$ and $\gamma_{eff} = 12.14$ m$^{-1}$ W$^{-1}$. For 3.54 eV pulse, the parameters are $\alpha = 11.2 \times 10^5$ m$^{-1}$, $\alpha_{2eff} = 5.64 \times 10^{-6}$ m W$^{-1}$, and $\gamma_{eff} = 0.0086$ m$^{-1}$ W$^{-1}$. The propagation distance for both the wavelengths is $Z = 1.1 \times 10^{-6}$ m. The transmission spectrum of metal-dielectric photonic crystal along with the energy of interest, are shown at the top inset for reference.

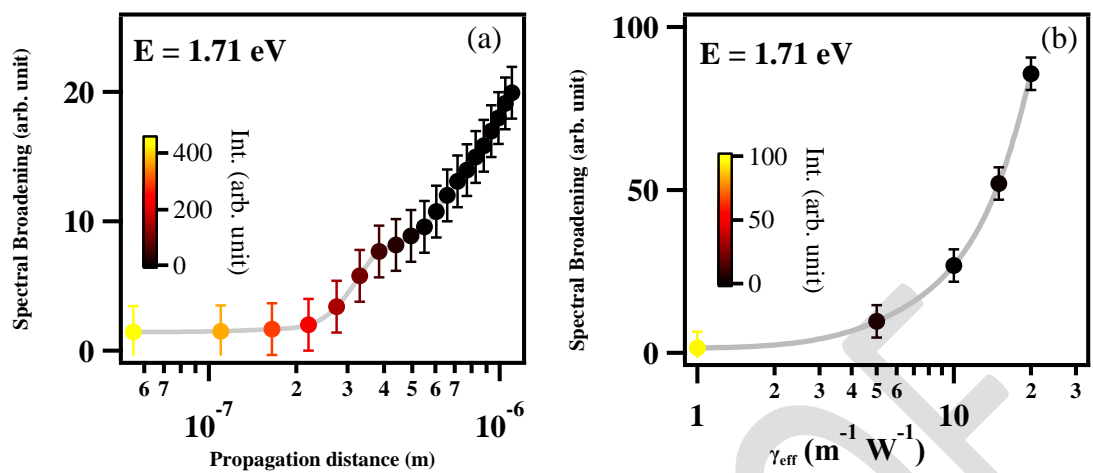

**Figure S7.** Spectral broadening during the pulse propagation **(a)** along with the propagation distance (Z) and **(b)** with varied Kerr nonlinearity ($\gamma_{eff}$). The spectral data is from Fig 6b and 6d, respectively.

## 6. Schematic experimental arrangements for optical pump-probe spectroscopy

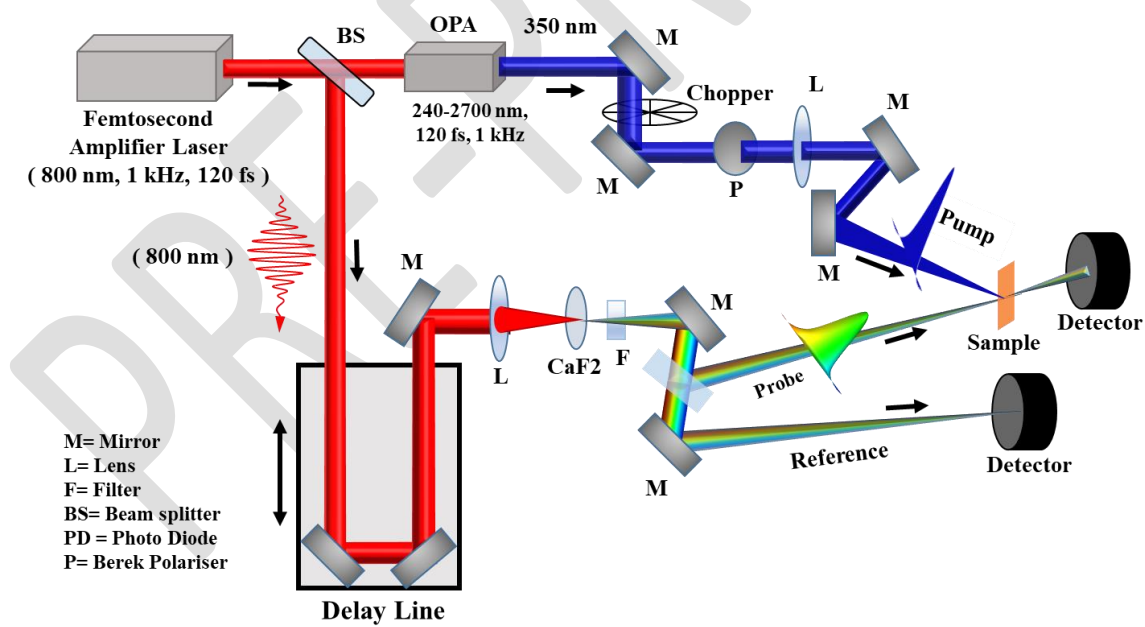

**Figure S8.** Schematic representation of femtosecond optical pump-probe experimental setups to perform the transient absorption spectroscopy.

_____